\documentclass[10pt]{article}

\usepackage{amsmath}
\usepackage{amssymb}
\usepackage{graphicx}
\usepackage{cite}
\usepackage{color}

\usepackage{setspace}
\usepackage[labelfont=bf,labelsep=period,justification=raggedright]{caption}

\makeatletter
\renewcommand{\@biblabel}[1]{\quad#1.}
\makeatother

\date{}

\begin{document}

\begin{flushleft}
{\Large
\textbf{Measuring multiple evolution mechanisms of complex networks}
}
\\
Qian-Ming Zhang$^{1,2,4}$
Xiao-Ke Xu$^{3,\ddag}$
Yu-Xiao Zhu$^{1,4}$
Tao Zhou$^{1,4,\S}$
\\
\bf{1} Web Sciences Center, School of Computer Science and Engineering,
University of Electronic Science and Technology of China, Chengdu 611731, People's Republic of China
\\
\bf{2} Center for Polymer Studies, Department of Physics, Boston University, Boston 02215, United States of America
\\
\bf{3} College of Information and Communication Engineering, Dalian Nationalities University, Dalian 116600, People's Republic of China
\\
\bf{4} Big Data Research Center, University of Electronic Science and Technology of China, Chengdu 611731, People's Republic of China
\\
Corresponding to: $\ddag$ xiaokeeie@gmail.com, $\S$ zhutou@ustc.edu
\end{flushleft}

\section*{Abstract}
Numerous concise models such as preferential attachment have been put forward to reveal the evolution mechanisms of real-world networks, which show that real-world networks are usually jointly driven by a hybrid mechanism of multiplex features instead of a single pure mechanism. To get an accurate simulation for real networks, some researchers proposed a few hybrid models of mixing multiple evolution mechanisms. Nevertheless, how a hybrid mechanism of multiplex features jointly influence the network evolution is not very clear. In this study, we introduce two methods (link prediction and likelihood analysis) to measure multiple evolution mechanisms of complex networks. Through tremendous experiments on artificial networks, which can be controlled to follow multiple mechanisms with different weights, we find the method based on likelihood analysis performs much better and gives very accurate estimations. At last, we apply this method to some real-world networks which are from different domains (including technology networks and social networks) and different countries (e.g., USA and China), to see how popularity and clustering co-evolve. We find most of them are affected by both popularity and clustering, but with quite different weights.


\section*{Introduction}
Many social, technological networks evolve over time after they are established. Previous studies have revealed that real networks possess many different structural features, like various degree distribution \cite{Amaral2000}, different levels of clustering \cite{Cluster2004}, existent or nonexistent communities \cite{Community}, assortative or disassortative mixing pattern \cite{newman2002}, long or short average shortest distance, and so on, which attract much attention on building models to mimic the network evolution \cite{ModelSurvey0,ModelSurvey}. Meanwhile, the latent mechanisms are also fruitful such as the rich-get-richer \cite{BA1999}, the good-get-richer \cite{ZhouPlos}, the stability constrains \cite{Perotti2009}, homophily \cite{McPherson2001}, clustering \cite{Cluster} etc. However, using one pure mechanism is usually insufficient to depict real-world networks precisely because of those different aspects of features. Therefore, researchers mixed different mechanisms in order to get better simulation, like the mixture of clustering and preferential attachment \cite{Cluster,Holme2002}, popularity and randomness \cite{YCLai2002}, popularity and similarity \cite{Papadopoulos2012}, topology distance and geographical distance \cite{Zhoutao2007}, and so on. In all, networks are likely to be driven by multiple mechanisms, and we are inspired to raise a question: is it possible to measure the contribution of each mechanism in the network evolution?

The inchoate way to evaluate network model or underlying mechanism is based on the comparison between some selected structural features. It supposes a model is better than another one if its generated network is more close to the target network in terms of those selected features. But such method cannot be well validated since no one has the fair standard to select representative one from countless structural features. Without considering any specific structural feature, we had proposed a method based on likelihood analysis to fairly evaluate network models \cite{Wang2012}. Therein, we can calculate the appearing likelihood for each newly created link according to the model's mechanism, and then multiply them together to get the likelihood of the set of new links. For a group of models, the one giving the highest likelihood is considered to be the most suitable one. This method is inspired by the link prediction approach, which aims at estimating the likelihood of the existence of a link based on the observed links \cite{LP2011}. According to this definition, if the principle of a link prediction algorithm is consistent to the mechanism of a given network, this algorithm should provide accurate predictions. Therefore, one can also evaluate the latent mechanisms according to the prediction results of the corresponding link prediction algorithms \cite{Zhang2013,Cannistraci2013}. In this paper, we take the latter two methods into consideration because they are both free of any specific structural features. To our knowledge, the above methods have only been applied to judge which mechanism is better given a series of mechanisms, but have never been applied to measure the contributions of multiple mechanisms in network evolution.

The core idea of the above methods is to estimate the appearing likelihood of links, which inspires us to measure the contributions of multiple mechanisms by calculating the likelihood using all the mechanisms simultaneously. Therefore, we design a formula to re-calculate the likelihood for every link by assigning each mechanism an tunable parameter. The optimal group of parameters are the ones maximizing the likelihood of all links (likelihood analysis method) or the prediction accuracy (link prediction method). To testify the effectiveness, we produce numerous model networks which can be controlled to follow multiple mechanisms with different weights, such as popularity, clustering and randomness. Through comparing the estimated contributions with the known weights, we find both of the methods are effective to judge which mechanism is stronger. In particular the one based on likelihood analysis can give very accurate estimations. Further, we discuss the advantage of likelihood analysis method and the disadvantage of the link prediction method which leads to its failure. At last, we apply this method to different kinds of  real-world networks to see how popularity and clustering co-evolve in real complex networks. These networks are collected from different domains, including technology networks and social networks, and from different countries, e.g. USA and China. The results show that most of these networks evolve with both mechanisms but with quite different weights.

The main contributions are two folds. In the theoretical aspect, we clarify that the multiple mechanisms of complex systems can be measured in quantitative way, and provide a unified, efficient and extensible measurement method. In the aspect of specific conclusions, we find some interesting and different properties for real-life networks. For example, the clustering mechanism widely exists in any social networks, while in the platform mainly designed for social activities (Facebook and Flickr) the clustering effect is much stronger than in the platform where the primary demands of users are not social intercourse, such as to watch videos in Youtube and to read blogs in ScienceNet. In addition, we showed that the evolving mechanisms may remarkably change in time for some real networks (e.g., Internet), so the links associated with new nodes are created with different reasons by links between old nodes, which are usually ignored in known models, but in accordance with some experimental studies on Internet, such as \cite{Carmi2007} and \cite{ZhangGQ2008}.

\section*{Results}
\subsection*{Measurement methods}
Given two snapshots of an evolving network at time $t_1$ and $t_2$ ($t_1 < t_2$), denoted by $G(V,E)$ and $G'(V',E')$ respectively, where $V$ ($V'$) and $E$ ($E'$) are the sets of nodes and links respectively. The set of new links is $E_{new}=E'-E$. In the following we firstly introduce two previous methods of evaluating underlying mechanisms in network evolution, and then present how we measure contributions of multiple mechanisms.

One method is based on likelihood analysis \cite{Wang2012}, of which the key idea is to estimate the appearing likelihood for each new link by multiply the probabilities of selecting its two endpoints. For example, if the links are all randomly created, the likelihood of each link $(x,y)$ can be calculated by $l_{xy}=\frac{1}{N}\cdot\frac{1}{N}$ where $N$ is the number of nodes of the network. Then, we can get the likelihood for all the new links according to $\mathbb{L}=\prod_{(x,y)\in E_{\mathrm{new}}} l_{xy}$. For a group of models, we can calculate $\mathbb{L}$ for each of them, and the one with the highest likelihood $\mathbb{L}$ is considered to be the most suitable one.

The other method is based on link prediction \cite{Zhang2013,Cannistraci2013}. The link prediction index would assign a score, following some certain principle, to each non-observed links, including new links $E_{\mathrm{new}}$ and nonexistent links $E_{\mathrm{non}}$ ($E_{\mathrm{non}}=U-E-E_{new}$, where $U$ is the universal set containing all $|V|(|V|-1)/2$ links). Then we can rank these links in descending order. A link prediction index is good if it can assign the new links higher rankings compared with the nonexistent links. To measure it in a quantified way, we introduce the AUC value (area under the receiver operating characteristic curve  \cite{AUC1982,LP2011}) which will be discussed in detail in \textbf{Materials and Methods}. Then we assume that a mechanism is more suitable to depict the network evolution if the corresponding link prediction algorithm results in a higher AUC.

As described above, the key points are both to estimate the likelihoods of links. We are motivated to re-estimate the likelihood by considering all the mechanisms with tunable parameters (which must sum to 1) indicating their contributions. According to the probability theory, we define the likelihood of link $(x,y)$ as the expectation of the likelihoods for all the mechanisms, written as
\begin{equation}\label{Eq_likelihood}
  l_{xy} = \sum_{i=1}^m\lambda_il_{xy}^{(i)} \quad, \qquad  (k>0)
\end{equation}
where $m$ is the number of considered mechanisms. Thus, for the method based on likelihood analysis, we expect the group of parameters which maximize $\prod_{(x,y)\in E_{\mathrm{new}}} l_{xy}$ would indicate the contribution of each mechanism. Similarly, for the method based on link prediction model, the group of parameters which maximize the prediction accuracy (AUC) would indicate the contribution of each mechanism.

\subsection*{Comparisons between the two methods}
To examine the effectiveness of the measurement methods, we apply them to model networks of which the evolution can be controlled. Two well-known mechanisms, popularity and clustering, are firstly taken into consideration. Popularity denotes that the nodes with higher degree are more attractive, while clustering suggests that the links which can form more triangles is more preferred. The model network evolves beginning with a loop consisting of five nodes. It grows following two rules at each step:
\begin{enumerate}
  \item[(i)] add one new node with one new link which connects this new node to one old node;
  \item[(ii)] add $3$ links, but self-loops and multi-links are not accepted.
\end{enumerate}
Every new link is created following either popularity mechanism or clustering mechanism, which is controlled by a tunable parameter $p$ ranging from 0 to 1. $p=0$ means all the links are created following popularity mechanism, while $p=1$ means all the links are created following clustering mechanism.

To implement popularity mechanism, we choose \emph{preferential attachment} which was depicted by Barab\'asi and Albert in \cite{BA1999}. They defined the probability of selecting node $x$ for new links as $\frac{k_x}{\sum_{z\in V}{k_z}}$. Similarly, for clustering mechanism we use the number of common neighbors to measure the likelihood of creating a link between $x$ and $y$. In detail, we firstly select a node $x$ for the new link, and then select the other node preferentially according to the probability $\frac{|\Gamma{(x)}\cap\Gamma{(y)}|}{\sum_{z\neq x}{|\Gamma{(x)}\cap\Gamma{(z)}|}}$, where $\Gamma{(x)}$ is the set of neighbors of $x$. Node $x$ is selected randomly to differ from popularity mechanism. Notice that, the new link which is added with the new node at each step, cannot be created if following the current clustering mechanism. So we randomly select an old node to form this new link to differ from \emph{preferential attachment}. By tuning $p$ from 0 to 1 with step-length 0.1, we respectively produce 100 model networks for every $p$. Then the question can be simplified to estimate the value $p$ for each model network through equation (\ref{Eq_likelihood}).

\textbf{Link prediction method.} Corresponding to the implementation of popularity mechanism, there has been proposed a link prediction index named Preferential Attachment (PA) index which is defined as the product of the degrees of two nodes, written as $s_{xy}^{\mathrm{PA}}=k_x\times k_y$ \cite{BA1999,LP2003,LP2011}. There also has been proposed Common Neighbor (CN) index \cite{LP2003} which is accordant with the clustering mechanism, written as $s_{xy}^{\mathrm{CN}}=|\Gamma{(x)}\cap\Gamma{(y)}|$.  Notice that, many node pairs have the same number of common neighbors, or no common neighbor, which leads to the indistinguishable $s_{xy}^{\mathrm{CN}}$ and the degeneracy of states \cite{Zhou2009}. To tackle such problems but keeping the predictive power of CN index invariant, we add a small random number $\varepsilon\in(0,0.01)$ to every $s_{xy}^{\mathrm{CN}}$, rewritten as $s_{xy}^{\mathrm{CN}}=|\Gamma{(x)}\cap\Gamma{(y)}|+\varepsilon$. Because $s_{xy}^{\mathrm{PA}}=k_x\times k_y$ is much larger than $s_{xy}^{\mathrm{CN}}\leq \mathrm{min}(k_x, k_y)$, we must normalize the $s_{xy}^{\mathrm{PA}}$ and $s_{xy}^{\mathrm{CN}}$ when we combine them. Otherwise we need larger $\lambda$ to strengthen the role of  $s_{xy}^{\mathrm{CN}}$ to keep away from bad estimation. Thus we define the hybrid index as
\begin{equation} \label{Eq_linkpred}
s_{xy}=(1-\lambda)\widehat{s_{xy}^{\mathrm{PA}}}+\lambda \widehat{s_{xy}^{\mathrm{CN}}},
\end{equation}
where $\widehat{s_{xy}^{\mathrm{PA}}}$ and $\widehat{s_{xy}^{\mathrm{CN}}}$ are the normalized values by the mean $s_{xy}^{\mathrm{PA}}$ and $s_{xy}^{\mathrm{CN}}$ respectively. In detail, $\widehat{s_{xy}^{\mathrm{PA}}} = s_{xy}^{\mathrm{PA}} / \langle{s_{xy}^{\mathrm{PA}}}\rangle$, and $\widehat{s_{xy}^{\mathrm{CN}}} = s_{xy}^{\mathrm{CN}} / \langle{s_{xy}^{\mathrm{CN}}}\rangle$, where $\langle \bullet\rangle$ is the mean value of $\bullet$. By tuning $\lambda$ ranging from $\in[0,1]$, we can easily find the optimal $\lambda$ which maximizes the prediction accuracy (AUC). Need to notice that, CN index can not work if any endpoint of a new link appears after $t_1$. So we remove all the new links with such nodes when to implement the link prediction method. To keep unanimous, such new links are also ignored when applying the likelihood analysis method.

\textbf{Likelihood analysis method.} This method \cite{Wang2012} defines the likelihood of a link $(x,y)$ as the multiplication of the likelihoods of selecting node $x$ and $y$. Thus, $l_{xy}^{\mathrm{popu}}$ can be easily defined as $\frac{k_x}{\sum{k_i}} \times \frac{k_y}{\sum{k_i}}$, and $l_{xy}^{\mathrm{clus}}$ can be defined as
$\frac{1}{2}(\frac{1}{N} \times \frac{|\Gamma{(x)}\cap\Gamma{(y)}|}{\sum_{z\neq x}{|\Gamma{(x)}\cap\Gamma{(z)}|}} +
\frac{1}{N} \times \frac{|\Gamma{(y)}\cap\Gamma{(x)}|}{\sum_{z\neq y}{|\Gamma{(y)}\cap\Gamma{(z)}|}} )$. Then the likelihood of $(x,y)$ has the format
\begin{equation} \label{Eq_likeliAnaly1}
l_{xy}=(1-\lambda)l_{xy}^{\mathrm{popu}} + \lambda l_{xy}^{\mathrm{clus}}.
\end{equation}
This model aims to maximize the likelihood of all the new links, written as
\begin{equation} \label{Eq_likeliAnaly2}
\mathbb{L}= \prod_{(x,y)\in E_{\mathrm{new}}}l_{xy}.
\end{equation}
Thus, we can also obtain the optimal $\lambda$ which maximizes $\mathbb{L}$. Notice that if $|\Gamma{(x)}\cap\Gamma{(y)}|=0$, $l_{xy}^{\mathrm{CN}}$ will be meaningless. Please see the solution in \textbf{Materials and Methods}, where we also define $l_{xy}$ if we consider new links without the limitation of new nodes.

\begin{figure} [!ht]
  \begin{center}
  \includegraphics[width=15cm]{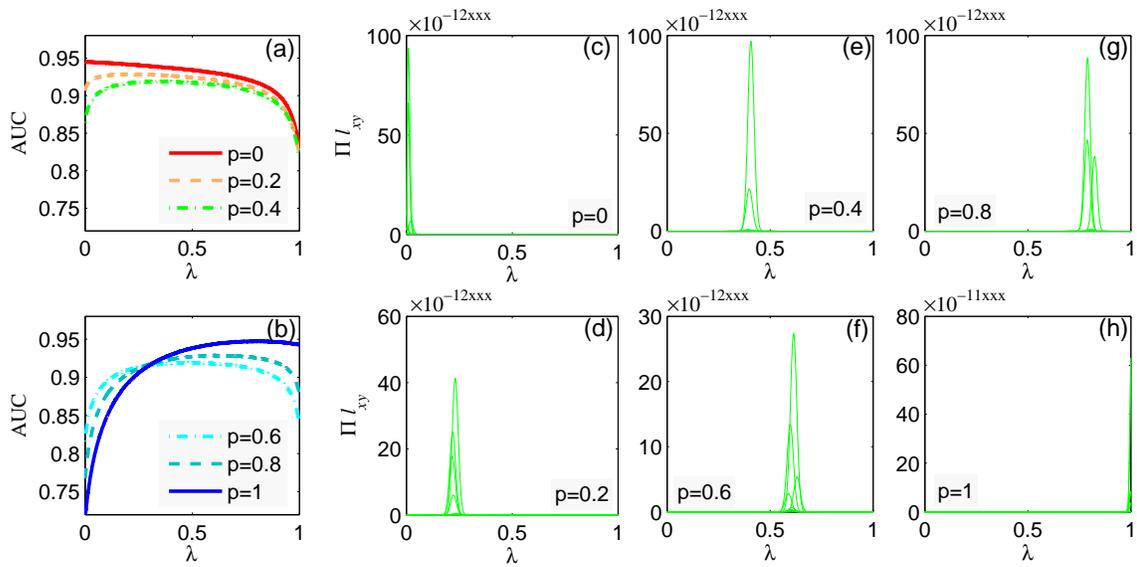}\\
  \end{center}
  \caption{\label{fig_trendLPandLI}
  Measuring popularity and clustering based on link prediction model and likelihood analysis respectively. The contributions are estimated through the peak values. Subfigure (a) and (b) present the average values of AUC resulted by link prediction model, which are obtained by averaging 100 implementations through 100 model networks. The others present the values of $\mathbb{L}$ resulted from likelihood analysis. Therein, each curve corresponds to one model network. $\lambda$ corresponds to the coefficient in equation (\ref{Eq_likelihood}). $p$ denotes the contribution of clustering mechanism in the model networks. Because the likelihoods for the networks are not in the same order of magnitude, we use 12xxx instead of the exact values. 12xxx means an uncertain value above 11999 and below 13000.}

\end{figure}

\begin{figure*} [!ht]
  \begin{center}
  \includegraphics[width=12cm]{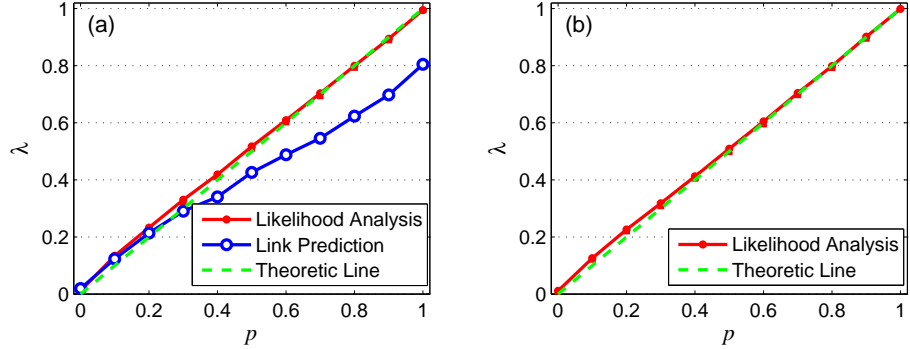}
  \end{center}
  \caption{\label{fig_Figmodel}
   Correlation between the optimal $\lambda$ and $p$. $p$ is the known proportion of clustering mechanism compared to popularity mechanism. $\lambda$ is the estimated value by the measurement method in this paper. Subfigure (a) represents the comparison between link prediction method and likelihood analysis, where no new links with new nodes are considered. Subfigure (b) only shows the results of likelihood analysis without the limitation of new nodes.}
\end{figure*}

In Figure \ref{fig_trendLPandLI}, we present the trends of AUC values (subfigure (a) and (b)) and $\mathbb{L}$ (subfigure (c)-(h)) with the increasing $\lambda$. The contributions of popularity mechanism and clustering mechanism can be estimated through the peak values. We can see that the optimal $\lambda$ resulted from both the two methods increase when $p$ grows bigger. For intuitive observation, we figure out the correlation between $p$ and the optimal $\lambda$ in Figure \ref{fig_Figmodel} (a). The likelihood analysis method gives very accurate estimation while the link prediction method fails when $p$ is large. The reasons of such failure are three folds: (i) CN mechanism embodies the principle of preferential attachment to some extent; (ii) the link prediction method provides too rough descriptions for the links; (iii) the link prediction model is not appropriate to measure the mechanisms' contributions.

Firstly, CN mechanism embodies the principle of preferential attachment because two nodes with large degrees have higher chance to have common neighbors. However, PA never considers the number of common neighbors shared by any node pair. When $p$ is small, few new links are restricted to form triangles. It's easy to distinguish CN mechanism from PA mechanism because most new links shares few, even no common neighbors. When $p$ becomes larger, although the formation of triangles become popular, the new links with many common neighbors also tend to have high-degree endpoints. There also exist many new links with few common neighbors but high-degree endpoints. These links lead to the failure of the link prediction method. We will explain it in detail through an example along with the third reason. However, this problem caused by the network model restricts the link prediction method but does not influence the likelihood analysis method. That should be due to the advantages of the likelihood analysis method, which are discussed as below.

The second reason is the loser's rough descriptions of the links compared with the winner. For example, suppose there are two pairs of unconnected nodes $(x,y)$ and $(u,v)$, which both have two common neighbors, but the degrees of $x$ and $y$ are much higher than those of $u$ and $v$. The probabilities that these links appear is obviously quite different, but the CN index assigns them the same values, i.e., $s_{xy}^{\mathrm{CN}}=s_{uv}^{\mathrm{CN}}$. In contrast, the likelihood analysis method can strongly distinguish them by applying probabilistic methods. Following the definition, we can get the likelihoods,
\begin{eqnarray}\label{Eq_2ndReason}
  l_{xy}&=&\frac{1}{2}(\frac{1}{N}\times \frac{2}{\sum_{z\neq x}|\Gamma(x)\cap\Gamma(y)|}+\frac{1}{N}\times \frac{2}{\sum_{z\neq y}|\Gamma(x)\cap\Gamma(y)|})   \nonumber\\
   &= &\frac{1}{N}(\frac{1}{\sum_{z\in\Gamma(x)}(k_z-1)} + \frac{1}{\sum_{z\in\Gamma(y)}(k_z-1)})
\end{eqnarray}
and $l_{uv}$ in the similar form, which are proved in \textbf{Materials and Methods}. Obviously, $l_{xy}$ is far different from $l_{uv}$, because $|\Gamma(x)|$ and $|\Gamma(y)|$ are much larger than both $|\Gamma(u)|$ and $|\Gamma(v)|$.

\begin{figure*} [!ht]
  \begin{center}
  \includegraphics[width=13cm]{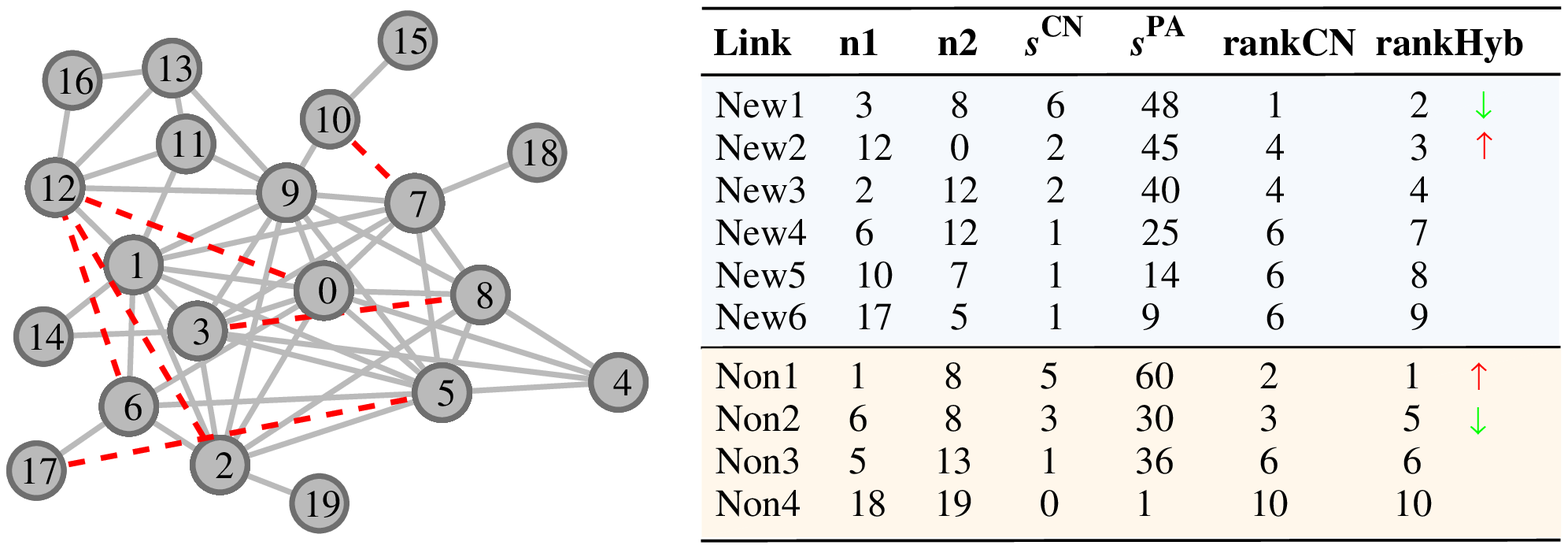}
  \end{center}
  \caption{\label{fig_examFailure}
  Example network driven by clustering mechanism only, and comparisons between the new links and some selected nonexistent links. Red dash links represent new links which are created following the clustering mechanism. New$i$ represents the IDs of new links, while Non$i$ represents the IDs of nonexistent links. The two end nodes of the link are labeled as $n1$ and $n2$. $s^\mathrm{CN}$ is the number of common neighbors between $n1$ and $n2$, corresponding to Common Neighbor Index. $s_\mathrm{PA}$ is calculated through Preferential Attachment Index. The numbers in ``rankCN'' column are the rankings based on $s_\mathrm{CN}$ (corresponding to $\lambda=1$), while those in ``rankHyb'' column are the rankings based on $0.8s^\mathrm{CN}+0.2s^\mathrm{PA}$(corresponding to $\lambda=0.8$).}
\end{figure*}

At last, in link prediction method, each new link needs to be compared with all the (sampled) nonexistent links. So that we can find the best link prediction index which assigns the new links with higher rankings compared with those nonexistent links. But when we try to improve the new links' rankings by tuning $\lambda$, there always exist some links whose rankings fall because of the improved rankings of some nonexistent links. That is to say, the nonexistent links, which are indispensable in the link prediction model, become the barriers to measuring the mechanisms' contributions. By comparison, the likelihood analysis method aims to optimize the overall likelihood of the new links as a whole. Until now, many researches discussed that some properties only emerge at the global level but vanish at the individual level, such as the function of the organs, the power-law distribution of displacement on the group level but not on the individual level \cite{Yan2013}, and so on. In our case, although the new links are created following CN mechanism when $p=1$, some of them might seem to be following PA mechanism as they have high-degree endpoints. Unless we consider the overall likelihood of these links, we cannot obtain the accurate estimation. Moreover, this method shakes off the effect of the nonexistent links. In fact, many pairs of unconnected nodes are deemed to be linked with high probability. These pairs of nodes would lead us astray if they are treated as the reference standard in the link prediction method. For clarity, we generate a small network following CN mechanism to explain such failure. As shown in Figure \ref{fig_examFailure}, new links are marked by red dash lines and New$i$. We also select six nonexistent links marked by Non$i$ to make comparisons. Clearly we can see that the node pair with high $s^{\mathrm{CN}}$ usually has high $s^{\mathrm{PA}}$, which is caused by the embodied preferential attachment principle. Such effect makes the estimation difficult. At first, we rank the links according to $s^{\mathrm{CN}}$, Non1 and Non2 are only behind New1. Then we introduce $s^\mathrm{PA}$, the rankings of New2 and New3 are improved due to their larger $s^\mathrm{PA}$, while Non2 with lower $s^\mathrm{PA}$ gets a lower ranking. Notice that, the prediction accuracy can benefit from such changes. However, we also need to notice the change happened on Non1, which will lower the accuracy. Non1 has both high $s^{\mathrm{CN}}$ and $s^{\mathrm{PA}}$ but belongs to nonexistent links. This is the ungovernable effect what we referred before. Adopting such link as the reference standard, it is difficult to obtain the accurate estimation.

As above, the likelihood analysis method wins due to its two advantages: the exact description of individual link, and the global perspective of description of all the new links. These two points are both indispensable. By comparison, the link prediction method is limited by its rough description of individual link, and the ungovernable effect of nonexistent links. To be more stringent, we redefine the CN index to get more accurate description of individual link by $s_{xy}^{\mathrm{CN'}}=\frac{1}{2}(\frac{1}{N} \times \frac{|\Gamma{(x)}\cap\Gamma{(y)}|}{\sum_{z\neq x}{|\Gamma{(x)}\cap\Gamma{(z)}|}} + \frac{1}{N} \times \frac{|\Gamma{(y)}\cap\Gamma{(x)}|}{\sum_{z\neq y}{|\Gamma{(y)}\cap\Gamma{(z)}|}} )$, which has the same form to the equation of the likelihood analysis method. But it still failed, as shown in Figure S1 in the Supporting Information. The result implies the effect of the nonexistent links is the main reason.

In Figure \ref{fig_Figmodel} (b), we show another advantage of the likelihood analysis method. Due to the drawback of link prediction model, we do not consider the new links with new nodes in Figure \ref{fig_Figmodel} (a), but such new links do not limit the effectiveness of likelihood analysis method. Actually, they can improve the accuracy of the estimation a little bit.

\subsection*{Verification through model networks with more mechanisms}
Without loss of generality, we examine the winner through model networks driven by more mechanisms. Thus we introduce randomness mechanism, which means that the endpoints of new links are all randomly selected. Similarly, the model networks start evolving from a loop consisting of five nodes. At each step, one new node with one new link and three other links are added. Every link is created following Randomness mechanism with probability $p_{\mathrm{rand}}$, clustering mechanism with probability $p_{\mathrm{clus}}$ or popularity mechanism with probability $p_{\mathrm{popu}}$, where $p_{\mathrm{rand}}, p_{\mathrm{clus}}, p_{\mathrm{popu}}\in[0,1]$, and $p_{\mathrm{rand}}+p_{\mathrm{clus}}+p_{\mathrm{popu}}=1$.

\begin{figure*} [!ht]
  \begin{center}
  \includegraphics[width=8cm]{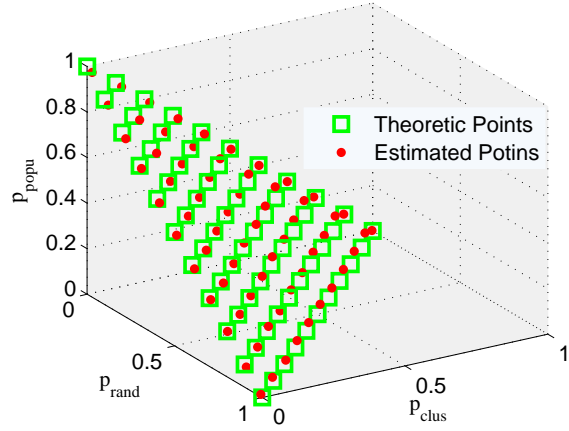}\\
  \end{center}
  \caption{\label{fig_FitmodelTri}
  The fitting degree of the estimated contribution and the theoretical values $p_{\mathrm{rand}}$, $p_{\mathrm{clus}}$ and $p_{\mathrm{popu}}$. Red spots denote the estimated values resulted from likelihood analysis method. Green rectangles mean the theoretical values.}
\end{figure*}

By calculating the $\mathbb{L}$ through equation (\ref{Eq_likeliAnaly2}), we can plot every group of estimated values $\{p_{\mathrm{rand}}$, $p_{\mathrm{clus}}$, $p_{\mathrm{popu}}\}$, in a three-dimensional space. As shown in Figure \ref{fig_FitmodelTri}, red spots denote the estimated values, while green rectangles show the locations of the theoretical values. The tight fitting again reflects the accurate estimation resulted by likelihood analysis method.

\subsection*{Measuring popularity and clustering for real networks}
Inspired by the effectiveness of the measurement method, we try to understand how popularity and clustering mechanism affect real-world networks. We collected nine networks including internet, social networks, communication networks and collaboration networks. Each of them is divided into two parts based on time stamps --- observed links and new links (see details in \textbf{Materials and Methods} and Table \ref{basicstat}).

\begin{figure*} [!ht]
  \begin{center}
  \includegraphics[width=12cm]{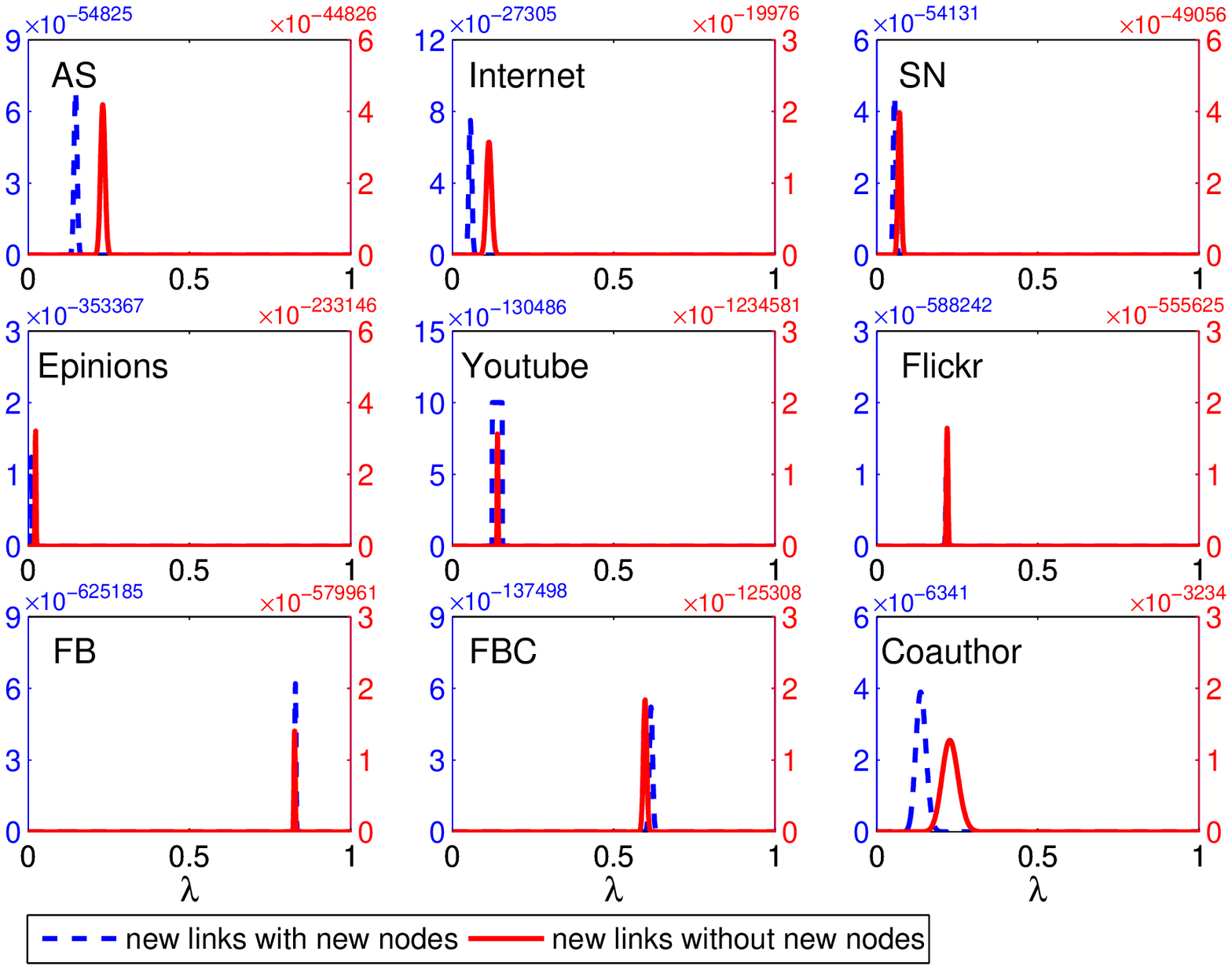}\\
  \end{center}
  \caption{\label{fig_Wangreal}
  The optimal $\lambda$ of likelihood analysis for real networks. Blue dash curves represent the likelihood calculated through new links without the limitation of new nodes, while red curves represent the likelihood calculated through new links without new nodes. }
\end{figure*}

By calculating the likelihood of new links with equation (\ref{Eq_likeliAnaly2}), we can also easily find the optimal $\lambda$ for every real network, indicated by the peaks of blue dash curves in Figure \ref{fig_Wangreal}. Obviously, the clustering mechanism widely exists in any social networks, but takes on different roles. The clustering effect is much stronger in the platform Facebook and Flickr, which are mainly designed for social activities where people tend to form clusters. Differently, in the platform of Youtube, ScienceNet and Epinions, the clustering effect loses to the popularity effect, because the primary demands of their users are not social intercourse but to watch videos (in Youtube), read blogs (in ScienceNet) and rate products (in Epinions). It does make sense because people who have better resources (e.g., excellent videos, great blogs) also hold greater appeal. In the collaboration network (Coauther), clustering and popularity also co-exist. In fact, many scientists have their own groups where advisors and students can collaborate with each other. However, the students usually leave after graduation and then replaced by new comers. The excellent students expect to join in the group held by successful scientists. In the next experiment, we can see that clustering effect would be a little stronger after they created the first link.

We further study the mechanisms for the new links among old nodes only, to observe the effect of new users. As shown by the red curves in Figure \ref{fig_Wangreal}, the optimal $\lambda$ tends to fall on different positions compared with the blue dash curves. The differences are not obvious in the online social platforms, but is significant in technology networks and collaboration networks. Such differences show that the evolving mechanisms may remarkably change in time, and the links associated with new nodes are created with different reasons by links between old nodes. This result on Internet is accordance with some previous experimental results \cite{Carmi2007,ZhangGQ2008}. Similarly, in the collaboration network, after a researcher joins a new group due to its reputation, he will develop more cooperations with other members.

\section*{Discussion}
Analyzing network evolution is not only a fundamental problem, but also a long-standing challenge in the network science domain. Previous studies focused on uncovering new mechanisms or improving some known mechanisms. In this paper, we started a new question that is to quantitatively measure the contributions of multiple mechanisms which affect the evolution of complex networks simultaneously. Motivated by previous studies, we compared two measurement methods which are based on link prediction and likelihood analysis respectively. Although the core ideas are both to estimate the likelihood for newly created links, the link prediction method fails in some cases. By analyzing their differences, we found the likelihood analysis method successfully captures the characteristics of new links on the individual level, and the overall property of new links on the group level as well. In fact, many researches have discussed that some features or functions emerge on the group level but vanish on the individual level, such as the function of the organs, the collective behaviors of the ant colonies, the power-law distribution of displacement on the group level but not on the individual level \cite{Yan2013}, etc. As a result the likelihood analysis method has the ability of producing very accurate estimations.

The likelihood analysis method is promising because it is highly extensible. The likelihood of new links can be easily estimated by counting the probabilities of choosing the two endpoints when given a mechanism. Moreover, this method is very efficient. Most of the computing time is consumed by the process of maximizing the likelihood, but this is a mature question in engineering. Therefore, it is possible to trace the evolution of complex systems in real time.

From the results of the real-world networks, we can clearly observe the combined action of popularity and clustering. The results here match our intuitive knowledge, but are more significant. For example, a network with high clustering coefficient \cite{Watts1998} is not necessarily driven by clustering mechanism, but probably the byproduct of another mechanism such as the spatially preferential attachment mechanism \cite{Marc2003}. Moreover, the value of clustering coefficient is usually dependent on the scale of networks, i.e., large scale networks usually have small clustering coefficient compared with small scale networks. None of the above cases can limit the likelihood analysis method, because the measurement of the links is directly based on the probability of selecting the endpoints following the given mechanism. In addition, we also showed that the evolving mechanisms may remarkably change in time for some real networks. Due to the efficiency of the likelihood analysis method, it is possible to trace the evolution of the networks and even the mechanisms. Our results suggests that the multiple mechanisms of complex networks can be measured in a quantitatively unified and efficient way. In future, we expect that the framework in this study can be used to provide some insights in understanding complex systems.

\section*{Materials and Methods}
\subsection*{Link Prediction Method}
Given $G(V,E)$, a link prediction index can assign every non-observed link (including $E_{new}$ and $E_{non}$) a score, according which we can rank these links in descending order. An index is regarded as better if it can order the links in $E_{new}$ with higher rankings than another index does. This is how we seek optimal $\lambda$ in this paper.

To compare the indices in a quantified way, we introduce AUC (area under the receiver operating characteristic curve \cite{AUC1982}) to measure the accuracy of prediction based on the rankings. It can be interpreted as the probability that a randomly chosen new link (a link in $E_\mathrm{new}$) is given a higher score than a randomly chosen nonexistent link. In the implementation, among $n$ times of independent comparisons, if there are $n'$ times the new link having higher score and $n''$ times the new link and the nonexistent link having the same score, we define the AUC value as \cite{LP2011}:
\begin{equation}
\mathrm{AUC} = \frac{n'+0.5n''}{n}.
\end{equation}
If all the scores are generated from an independent and identical distribution, the AUC value should be about 0.5. Therefore, the degree to which the AUC value exceeds 0.5 indicates how much better the algorithm performs than pure chance. Need to notice that, the calculation of AUC is based on statistical theory, so the result of equation (5) will be more approximate to the real value if we assign $n$ a larger number. We have discussed the proper value of $n$ in the book named Link Prediction \cite{LPBook2013}. That is, if we expect to get the AUC value with error less than 0.001 at the 90\% confidence level, $n$ should be no less than 672400. So in our experiments, we set $n=673000$. The derivation process is presented in \textbf{Supplementary Information}.

\subsection*{Likelihood Analysis Method}
In this method, we need to consider three cases for a chosen link $(x,y)$: (i) either $x$ or $y$ is a new node, which appears after $t_1$; (ii) both $x$ and $y$ are new nodes; (iii) both of them are old nodes.

For popularity mechanism, if one of them is new node, supposed as $x$, then $l_{xy}^{\mathrm{popu}}=1\times\frac{k_y}{\sum{k_z}}$, where $z\in V$. If both of them are new nodes, $l_{xy}^{\mathrm{popu}}=1$. And if both of them are old nodes, $l_{xy}^{\mathrm{popu}}=\frac{k_x}{\sum{k_z}}\times\frac{k_y}{\sum{k_z}}.$

For clustering mechanism, once $x$ or/and $y$ are new nodes, no common neighbor they would share. Then we define, according to the implementation of clustering mechanism, $l_{xy}^{\mathrm{clus}} = 1\times\frac{1}{N}$ if one of them is new node, and $l_{xy}^{\mathrm{clus}} = 1$ if both of them are new nodes. If both of $x$ and $y$ are old nodes, $l_{xy}^{\mathrm{clus}} =
\frac{1}{2}(\frac{1}{N} \times \frac{|\Gamma{(x)}\cap\Gamma{(y)}|}{\sum_{z\neq x}{|\Gamma{(x)}\cap\Gamma{(z)}|}} +
\frac{1}{N} \times \frac{|\Gamma{(y)}\cap\Gamma{(x)}|}{\sum_{z\neq y}{|\Gamma{(y)}\cap\Gamma{(z)}|}} )$. Denote that, if $x$ and $y$ do not share any common neighbors, $l_{xy}^{\mathrm{clus}}$ here need be modified to keep $\mathbb{L}$ away from 0. In such case, we re-define $l_{xy}^{\mathrm{clus}} = \frac{1}{\sum{k_z}}\times\frac{1}{N}$ due to two reasons: (i) $l_{xy}^{\mathrm{cluster}}$ can not be 0, or else the product will be 0 too; (ii) $l_{xy}^{\mathrm{cluster}}$ must be small and may be variant for different networks. So we adopt the certain value which is not more than the probability of select one node following popularity mechanism.

For randomness mechanism, if one of $x$ and $y$ is new node, $l_{xy}^{\mathrm{rand}} = 1\times\frac{1}{N}$. If both of them are new nodes, $l_{xy}^{\mathrm{clus}} = 1$. And if both of them are old nodes, $l_{xy}^{\mathrm{clus}} = \frac{1}{N}\times\frac{1}{N}$.

\subsection*{Proof of Equation (\ref{Eq_2ndReason})}
The proof of $l_{xy}=\frac{1}{2}(\frac{1}{N}\times \frac{2}{\sum_{z\neq x}|\Gamma(x)\cap\Gamma(z)|}+\frac{1}{N}\times \frac{2}{\sum_{z\neq y}|\Gamma(y)\cap\Gamma(z)|}) = \frac{1}{N}(\frac{1}{\sum_{z\in\Gamma(x)}(k_z-1)} + \frac{1}{\sum_{z\in\Gamma(y)}(k_z-1)})$ can be reduced to proving $\sum_{z\neq x}|\Gamma(x)\cap\Gamma(z)| = \sum_{z\in\Gamma(x)}(k_z-1)$. The number of common neighbors between $x$ and $z$ is equal to the number of the 2-steps paths, denoted as $\sum_{u}\mathbb{P}(x,u,z)$, where $\mathbb{P}(x,u,z)=1$ if the path $(x,u,z)$ exists, namely $u$ is the common neighbor of $x$ and $z$; otherwise $\mathbb{P}(x,u,z)=0$. Then $\sum_{z\neq x}|\Gamma(x)\cap\Gamma(z)| = \sum_{z\neq x}\sum_{u}\mathbb{P}(x,u,z) = \sum_u\sum_{z\neq x}\mathbb{P}(x,u,z)$. Given the nodes $x$ and $u$, $\sum_{z\neq x}\mathbb{P}(x,u,z)$ can be considered as the amount of the 2-steps paths ($x,u,z$). That is to say, both $x$ and $z$ must be the neighbors of $u$. Therefore, the amount of the 2-steps paths is equal to $|\Gamma(u)-x|$ because $z\neq x$, namely $\sum_{z\neq x}\mathbb{P}(x,u,z)=k_u-1$. Moreover, $\mathbb{P}(x,u,z)=0$ if $u$ is not connected to $x$ directly, we can eventually prove that $\sum_{z\neq x}|\Gamma(x)\cap\Gamma(z)| = \sum_{u\in\Gamma(x)}\sum_{z\neq x}\mathbb{P}(x,u,z) = \sum_{u\in\Gamma(x)}(k_u-1) = \sum_{z\in\Gamma(x)}(k_z-1)$.

\subsection*{Data Description}
We collect nine networks and divide every one of them into two parts --- observed links and future links (corresponding to $E$ and $E_{new}$ respectively defined in the previous section), basing on the time-stamps. The basic features are listed in Table \ref{basicstat}.

(1) AS --- Autonomous system (AS) within Internet is a collection of connected Internet Protocol networks and routers under the control of one entity. Route-views Project collected the Internet at the AS level at many different times, and here we use the data of June 2006 to compose the Observed Links and that of December 2006 to compose the Future Links \cite{route,ZhangGQ2008}.

(2) Internet --- The Internet can be viewed as a collection of autonomous systems (AS) whose snapshots was created weekly by CAIDA (Center for Applied Internet Data Analysis). Mislove downloaded the entire history of their measurements which covered the period from January 5th, 2004 until July 9th, 2007 \cite{Mislove2009}. In this paper, we choose the date November 20th, 2006 as the watershed of Observed Links and Future Links so the size of future links can be approximated to 10\% of observed links.

(3) SN --- ScienceNet (\emph{www.sciencenet.cn}) is a virtual community for Chinese-speaking scientists. This data consisting of two snapshots --- July 22nd 2013 and August 12th 2013, is newly crawled from the web site by Xing Yu.

(4) Epinion --- Epinions (\emph{www.epinions.com}) is an online product rating site where users are connected by trust or distrust relationships. In the simplest case, we neglect the types of connections. The earliest link in the initial data \cite{Paolo2005} was collected on September 1st, 2001, while the latest was on August 11th, 2003.

(5) Youtube --- YouTube (\emph{www.youtube.com}) is a popular video-sharing site that also involves a social network. The initial data, consisting of links created before Jan. 15th 2007, was collected by Mislove \cite{Mislove2009}.

(6) Flickr --- Flickr (\emph{www.flickr.com}) is a photo-sharing site based on a social network. This data is collected by Mislove \emph{et al.} \cite{Mislove2008} and consisting of $2570535$ users and $33140018$ links in total. Here we only use a small sample by choosing out the links with time stamps 2006-11-02 and 2006-11-03. The links created at 2006-11-03 are considered as future links and the rest of links compose the observed network.

(7) FB --- Facebook (\emph{www.facebook.com}) is a social networking service and has over one billion users. The initial data in \cite{Viswanath2009} are crawled between January 20th, 2009 and January 22nd, 2009. The time of link establishment is signed by a UNIX time-stamp unless it can not be determined. We set all the undetermined time-stamps as 1.

(8) FBC --- This data is from \emph{www.facebook.com} but different from the friendships in FB. In this data, if a user $u$ post to another user $v$'s wall on Facebook, the directed link will be created from $u$ to $v$. Since users may write multiple posts on a wall or their own wall, the network collected in \cite{Viswanath2009} allowed multiple edges and loops. In this paper, we remove the loops and redundant edges (multiple edges which have appeared before).

(9) Coauthor --- This is a collaboration network from the e-print arXiv, which covers scientific collaborations between authors whose papers are submitted to High Energy Physics - Theory category. The data covers papers in the period from January 1993 to April 2003 \cite{Jure2007}. Notice that two authors may collaborate multi-times, which is simply represented by an unweighted link in this paper. The time-stamps are determined by their first collaboration.

\section*{Acknowledgments}
We acknowledge the useful discussion with Junming Huang and Wen-Qiang Wang. We also want to thank Xing Yu for collecting data. This work is jointly supported by the National Natural Science Foundation of China under Nos. 11222543 and 11205042. XKX was supported by the National Natural Science Foundation of China (Nos. 61004104, 61374170) and CCF-Tencent Open Research Fund (AGR20130112). QMZ and YXZ acknowledge the support from the Program of Outstanding PhD Candidate in Academic Research by UESTC (Nos. YBXSZC20131034 and YBXSZC20131035) and China Scholarship Council (No. 201306070064 and 201206070003).

\section*{Author Contributions}
Q.M., X.K. and T. designed the experiments. Q.M and Y.X implemented the experiments. Q.M. and X.K. interpreted the experimental findings. Q.M and T. wrote the main manuscript, which was revised by all authors.

\section*{Additional Information}
Competing financial interests: The authors declare no competing financial interests.

\section*{Figure Legends}

\section*{Tables}
\begin{table*}[!ht]
  \caption{\label{ex} The basic information of the real networks. $|V|$ is the number of nodes and $|E|$ is the number of links before $t_1$. $C$ and $r$ are clustering coefficient\cite{Watts1998} and assortative coefficient\cite{newman2002}, respectively. $\langle k\rangle$ is the average degree of network. $H$ denotes the degree heterogeneity defined as $H =\frac{\langle k^2\rangle}{\langle k\rangle^2}$. $|V_{\mathrm{new}}|=|V'-V|$ and $|E_{\mathrm{new}}|=|E'-E|$ are the numbers of new nodes and links during $(t_1,t_2)$. $|E_{\mathrm{new}}'|$ denotes the number of new links among old nodes only.}
  \centering
  {\begin{tabular}{l|cccccc|ccc}
  \hline
  \hline
  Networks &$|V|$ &$|E|$ & $C$ & $r$ & $\langle k\rangle$ & $H$ & $|V_{\mathrm{new}}|$ & $|E_{\mathrm{new}}|$& $|E_{\mathrm{new}}'|$\\
  \hline
  AS        & 22960     & 49545     & 0.354 & -0.196    & 4.32  & 62.34   & 2143 & 9723      &6346 \\
  Internet  & 23670     & 47079     & 0.334 & -0.202    & 3.98  & 64.63   & 1856 & 5333      &2824 \\
  SN        & 39748     & 249685    & 0.271 & -0.163    & 12.56 & 33.44   & 692  & 8213      &6541\\
  Epinions  & 117719    & 640152    & 0.251 & -0.07     & 10.88 & 21.21   & 13861& 71058     &29548\\
  Youtube   & 1022090   & 2690294   & 0.177 & -0.033    & 5.26  & 90.03   & 116409& 300149   &122287\\
  Flickr    & 1486725   & 11786888  & 0.379 & -0.02     & 15.86 & 50.59   & 4060 & 64734     &57882 \\
  FB        & 59699     & 735380    & 0.25  & 0.181     & 24.64 & 3.47    & 4032 & 81710     &70850 \\
  FBC       & 43590     & 165070    & 0.130 & 0.22      & 7.57  & 3.15    & 2223 & 18342     &15249  \\
  Coauthor  & 10093     & 15432     & 0.704 & -0.017    & 3.06  & 4.66    & 838  & 1716      &459  \\
  \hline
  \hline
  \end{tabular}
  \label{basicstat}}
\end{table*}

%

\end{document}